\documentclass[conference]{IEEEtran}
\IEEEoverridecommandlockouts
\usepackage{cite}
\usepackage{amsmath,amssymb,amsfonts}
\usepackage{algorithmic}
\usepackage{eurosym}
\usepackage{graphicx}
\usepackage{textcomp}
\usepackage{xcolor}
\usepackage{tikz}
\def\BibTeX{{\rm B\kern-.05em{\sc i\kern-.025em b}\kern-.08em
    T\kern-.1667em\lower.7ex\hbox{E}\kern-.125emX}}
    
\usepackage{acronym}
\acrodef{res}[RES]{Renewable Energy Source}
\acrodef{rec}[REC]{Renewable Energy Community}
\acrodefplural{rec}[RECs]{Renewable Energy Communities}
\acrodef{pv}[PV]{Photovoltaic}
\acrodef{sh}[SH]{Smart Home}
\acrodef{pa}[PA]{Phases-based Appliance}
\acrodef{acs}[ACS]{Air-Conditioning System}
\acrodef{ul}[UL]{Uncontrollable Load}
\acrodef{pevrs}[PEV-RS]{Plug-in Electrical Vehicle Recharging Station}
\acrodef{pev}[PEV]{Plug-in Electrical Vehicle}
\acrodef{cop}[COP]{Coefficient of Performance}
\acrodefplural{cop}[COPs]{Coefficients of Performance}
\acrodef{cm}[CM]{Community Manager}
\acrodef{soc}[SoC]{State of Charge}
\acrodef{cma}[CMA]{Community Management Algorithm}
\acrodef{moa}[MOA]{Member Optimization Algorithm}
\acrodef{coa}[COA]{Community Optimization Algorithm}
\acrodef{mpc}[MPC]{Model Predictive Control}
\acrodef{der}[DER]{Distributed Energy Source}
\acrodef{se}[SE]{Shared Energy}

\usepackage{siunitx}
\DeclareSIUnit \voltampere { VA }
\DeclareSIUnit \wh { Wh }
\DeclareSIUnit \rad { rad }

\usepackage[ruled,vlined]{algorithm2e}
\usepackage{graphics}
\usepackage{multirow}

\newcommand\copyrighttext{%
\footnotesize
  \centering\copyright~2023 IEEE. Personal use of this material is permitted. Permission from IEEE must be obtained for all other uses, in any current or future media, including reprinting/republishing this material for advertising or promotional purposes, creating new collective works, for resale or redistribution to servers or lists, or reuse of any copyrighted component of this work in other works. \\ Presented at the 15th IEEE PowerTech 2023, doi: 10.1109/POWERTECH55446.2023.10202822.}
\newcommand\copyrightnotice{%
\begin{tikzpicture}[remember picture,overlay]
\node[anchor=south,yshift=0pt] at (current page.south) {\setlength{\fboxrule}{0pt}\fbox{\parbox{\dimexpr\textwidth-\fboxsep-\fboxrule\relax}{\copyrighttext}}};
\end{tikzpicture}%
}
    
\begin{document}

\title{Optimal Coordination and Discount Allocation\\ in Residential Renewable Energy Communities\\ with Smart Home Appliances 
\thanks{Funded from POR FESR Lazio 2014-2020 under project no. A0375-2020-36770.}
}

\author{\IEEEauthorblockN{F. Conte}
\IEEEauthorblockA{\textit{Faculty of Engineering} \\
\textit{Campus Bio-Medico University of Rome}\\
Roma, Italy \\
f.conte@unicampus.it}
\and
\IEEEauthorblockN{F. Silvestro, A. Vinci}
\IEEEauthorblockA{\textit{DITEN} \\
\textit{University of Genova}\\
Genova, Italy \\
federico.silvestro@unige.it}
\and
\IEEEauthorblockN{A. R. Di Fazio}
\IEEEauthorblockA{\textit{DIEI} \\
\textit{University of Cassino and Southern Lazio}\\
Cassino (FR), Italy \\
a.difazio@unicas.it}
}

\IEEEaftertitletext{\copyrightnotice\vspace{1.1\baselineskip}}
\maketitle

\begin{abstract}
This paper proposes an optimal management strategy for a Renewable Energy Community defined according to the Italian legislation. The specific case study is composed by a set of houses equipped with smart appliances, that share a PV plant. The objective is to minimize the cost of electrical energy use for each member of the community, taking into account the discount achievable from government incentives with proper shaping of the community daily consumption. Such incentives are indeed proportional to the shared energy, \textit{i.e.} the portion of the renewable energy consumed at each hour by community members. The management algorithm allows an optimal coordination of houses power demands, according to the degree of flexibility granted by users. Moreover, a policy to fairly distribute the obtained discount is introduced. Simulation results show the potentialities of the approach.
\end{abstract}


%%%%%%%%%%%%%%%%%%%%%%%%%%%%%%%%%%%%%%%%%%%%%%%%%%%%%%%%%%%%%
\section{Introduction}
%%%%%%%%%%%%%%%%%%%%%%%%%%%%%%%%%%%%%%%%%%%%%%%%%%%%%%%%%%%%%
One of the key challenges to complete the worldwide energy transition process is the integration of \acp{res} in the power systems. 
Indeed, \ac{res} are characterized by a non-programmable nature and, often, are \acp{der}, \textit{i.e.} connected to the MV/LV networks and installed together with commercial and residential loads. Consequently, they have a strong impact on the operation of both distribution and transmission systems, making more complex their management and control.

Research is widely engaged in developing innovative solutions to enable the integration of \acp{der} into the power system. One of the most promising solutions, identified also by government institutions, are \acp{rec}. In Europe, they have been formally introduced with Regulation 2018/2001/EU \cite{REDII} as associations of citizens, commercial activities, enterprises, and local authorities that own small-scale \ac{res} power plants. In Italy, the EU Regulation has been transposed into law between 2020 and 2021 (Law 8/2020 and Legislative Decree 199/2021 - RED II). 

A specific feature of \acp{rec} is the type of connection among members. In Italy, such a connection is not physical but \textit{virtual}, i.e., each \acp{rec}' member is physically connected to the utility grid and has a contract with an electricity retailer. The whole community receives an incentive by the government on the basis of the \ac{se}, defined as the quantity of renewable energy produced and immediately used by the \ac{rec} members (i.e., on hourly basis). The incentive is then transformed into an extra-benefit for prosumers and a discount on the bill for consumers belonging to the \ac{rec}. 

Therefore, minimizing the cost of the bill while maximising extra-benefit are the main objectives of the community members. To carry out such an optimal operation, the \ac{rec} should implement suitably designed methodological and technological tools. In general, assuming a certain level of flexibility granted by the members in controlling their power generation and demand profiles, the maximization of the \ac{se} requires a coordination among the members decisions.

Literature provides many solutions to optimally operate a \ac{rec}. In \cite{Faia:2021} and \cite{Pereira:2022}, peer-to-peer trading strategies among community members that act as prosumers are proposed. In \cite{Feng:2020,Safdarian:2021}, coalitional game-based methods are proposed to optimally manage a \ac{rec} that interfaces with the energy market as a unique entity. In \cite{VanCutsem:2020}, a fully decentralized cooperative energy market is defined into a community of smart buildings using block-chain smart contracts. In \cite{Hafiz:2019}, a scenario-based stochastic optimization is adopted to minimize the cost of the electrical energy of a set of residential loads that share a \ac{pv} plant connected to a common DC bus. In \cite{DiLorenzo:2021} \ac{mpc} is adopted to minimize the energy use cost in a condominium equipped with a common \ac{pv} plant; in this work, each user remains passive towards the distributor except for a single active user that assumes the role of balancing node.

Although research is widely active in the study of solutions for the optimal management of \acp{rec}, to the best of our knowledge, there are still few works on case studies that comply with the Italian regulatory framework. Most of them are mainly dedicated to optimal sizing and planning of \acp{rec} \cite{Zatti:2021,Moncecchi:2020,Ghiani:2019,Galici:2021}. 

In \cite{Conte:2022a,Conte:2022b,DiFazio:2022} the \ac{se} is maximized according to the Italian law definition. In these works, \ac{mpc} is applied to some case studies where a common \ac{pv} plant is coupled with an energy storage system so as to maximize the \ac{se}, leaving the members free to consume electrical energy at their convenience. 

In this paper, extending and improving the approach outlined in \cite{Conte:2022a,Conte:2022b,DiFazio:2022}, a different case study is considered. The \ac{rec} is composed by a set of \acp{sh} that share a unique \ac{pv} plant. Houses are equipped with smart appliances, that can be controlled remotely by a \ac{cm}. The latter has the objective to optimally distribute members' consumption along the day taking perceiving the minimization of their bills, discounted proportionally to the \ac{se}. Home users (\textit{i.e.} the \ac{rec} members) are only required to indicate their preferences about the use of the smart appliances, but without any advance. The \ac{cm} manages the community adopting mixed-integer \ac{mpc} \cite{Morari:1988}. Moreover, a policy to fairly distribute the obtained discount is discussed and implemented through a second \ac{mpc}-based method.

The rest of the paper is organized as follows. In Section~\ref{sec:rec} the Italian regulatory framework on \acp{rec} is detailed. Section~\ref{sec:scenario} introduces the case study and discuss the discount allocation policy. Section~\ref{sec:algorithm} describes the community optimal management algorithm. Simulation results are provided in Section~\ref{sec:results}. Finally, conclusions are summarized in Section~\ref{sec:conclusions}.

%%%%%%%%%%%%%%%%%%%%%%%%%%%%%%%%%%%%%%%%%%%%%%%%%%%%%%%%%%%%%%
\section{REC in Italy}\label{sec:rec}
%%%%%%%%%%%%%%%%%%%%%%%%%%%%%%%%%%%%%%%%%%%%%%%%%%%%%%%%%%%%%%
In Italy, \acp{rec} are defined and regulated by the Law 8/2020 and the Legislative Decree 199/2021 - RED II. The main rules are detailed in the following.
\begin{itemize}
    \item A \ac{rec} is a legal entity, based on the open and voluntary participation, autonomous and effectively controlled by its members.
     \item Selling energy cannot be the main business activity of \ac{rec} members; its primary purpose is to provide environmental, economic or social community benefits.
     \item \ac{rec} members must be connected under the same primary HV/MV substation.
     \item Some \ac{rec} members must invest (singularly or together) in  new shared \ac{res} power plants (one or more), each one with a rated power lower than 1 MVA.
     \item Each \ac{rec} member has its own contract with any electrical energy retailer on the base of which it pays a bill. 
\end{itemize}

A peculiar characteristic of the Italian \ac{rec} is the virtual configuration of the community: the users are connected to their Points Of Delivery (PODs) and the energy is shared using the public distribution grid. In such a way, development of \acp{rec} does not require any physical change in the structure of the network.

In Italy, the collective self-consumption is promoted by defining the \ac{se} as the quantity of renewable energy produced and used 
by prosumers belonging to the ac{rec}; it is evaluated in an hour basis. Mathematically, \ac{se} is defined as the minimum between the delivered renewable energy and the aggregated consumption of the \ac{rec} members.

For the economic evaluation of the \ac{se}, each user must be equipped with smart meters that collect energy consumption and renewable production during the hours of the day. Based on these measurements, the \ac{rec} receives a payment consisting of three components:
\begin{itemize}
    \item[i.] an incentive (explicitly) proportional to the \ac{se}; 
    \item[ii.] the restitution of a part of the energy use tariff, according to a cost-reflective network charge remuneration logic (anyway proportional to the \ac{se});  
    \item[iii.] the remuneration of the energy delivered to the grid by each \ac{res} plant at the market price.
\end{itemize}
The first regulatory proposal assumed the sum of the incentive i. and the tariff restitution ii. equal to about $\gamma_{sh}$\,=\,0.10\,-\,0.11~\euro/kWh.  

It is worth remarking that the incentive obtained by the \ac{rec} cannot be considered as an economical profit, since \ac{rec} cannot act in the market as an energy producer. The incentive can instead be framed as a discount on the bills paid by individual members and a extra-benefit for prosumers. The Italian law does not indicate how this incentive should be divided among the \ac{rec} members. Indeed, they are free to establish its allocation according to a peer-to-peer private agreement.

%%%%%%%%%%%%%%%%%%%%%%%%%
\section{\ac{rec} Configuration and Discount \\ Allocation Policy}\label{sec:scenario}
%%%%%%%%%%%%%%%%%%%%%%%%%
\begin{figure}[t]
	\centering
    \includegraphics[width=0.8\columnwidth]{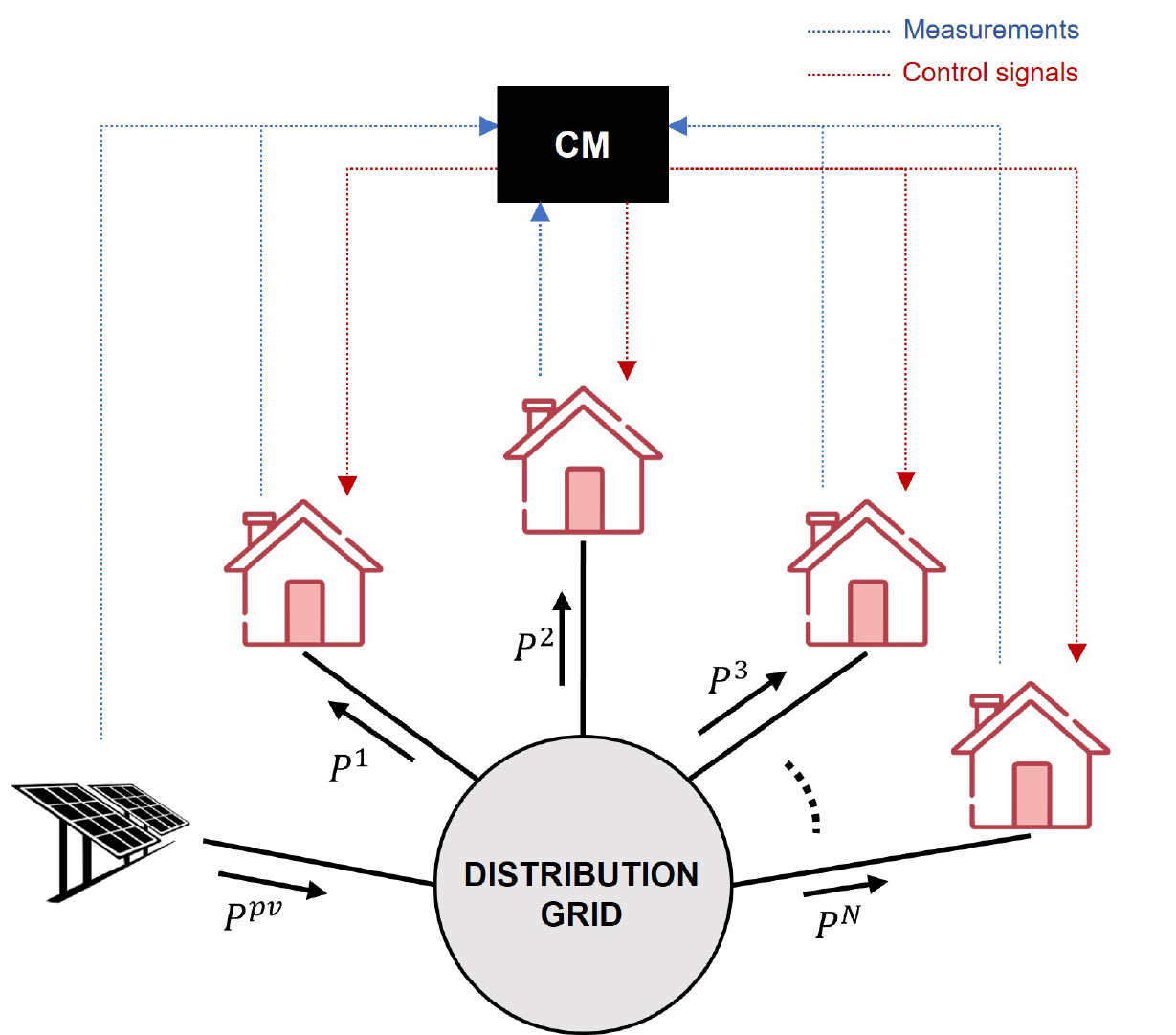}
\caption{\ac{rec} configuration.}
\label{fig:rec-configuration}
\end{figure}
In this paper, we consider the virtual configuration depicted in Fig.~\ref{fig:rec-configuration}. The \ac{rec} is composed by a common \ac{pv} plant, whose property is equally divided among the community members, and $N$ \acp{sh}. The operation of the \ac{rec} is monitored and controlled by the \ac{cm} at discrete time steps $k\Delta t_s$ with $k=0,1,2,\ldots$, where $\Delta t_s$ is the hourly sampling time interval.

Each \ac{sh} consumes in the considered time interval a power $P^i_k$ with $i=1,2,\ldots,N$. Analogously, the \ac{pv} plant generates a power $P^{pv}_k$. No common loads or energy storage systems are present in the \ac{rec}.

Each owner of \ac{sh} pays a different bill according to its contract with the electrical energy retailer. Specifically, the energy use cost for the $i$-th \ac{sh} at time step $k$ is 
\begin{equation}\label{eq:bills} 
    C_k^i = \gamma^i_k\cdot \Delta t_s P_k^i,
\end{equation}
where $\gamma_k^i$ is the time-varying energy tariff. It is worth remarking that the fixed part of bills are not considered, since they do not vary by modifying consumption profiles.    

As detailed in Section~\ref{sec:rec}, according to the Italian legislation the \ac{rec} obtains a common amount equal to
\begin{equation}\label{eq:discount}
    D_k =  \gamma_{sh} \Delta t_s P_k^{sh} + \gamma^e_k \Delta t_s P^{pv}_k,
\end{equation}
where $\gamma^e_k$ is the energy market price and $\Delta t_s P_k^{sh}$ is the \ac{se}, defined according to
\begin{equation}\label{eq:shared}
    P_k^{sh} = \min\left\lbrace P_k^{pv},\sum_{i=1}^N P_k^i\right\rbrace.
\end{equation}

Each \ac{sh} is equipped with a set of controllable and uncontrollable appliances (as detailed in the following Section~\ref{sec:models}). The \ac{cm} can remotely control these appliances to modify the \acp{sh}' power profiles $\{P_k^i\}_{k\geq 0}$. This action will change the energy use costs \eqref{eq:bills}, and, in turns, $P^{sh}_k$ and the first term in \eqref{eq:discount}. Thus, the objective of the \ac{cm} is to minimize the following cost function
\begin{equation}\label{eq:cer-cost-function}
    J = \Delta t_s \sum_{k\in \mathcal{K}} \left[\sum_{i=1}^N (\gamma^i_k P_k^i) -  \gamma_{sh} P_k^{sh}\right]
\end{equation}
over a given time interval $\mathcal{K}$ (\textit{e.g.}, one month) by varying the \acp{sh}' consumption.

After the optimization and the \ac{rec} operation, the obtained power profiles are indicated as $\{\hat{P}_k^i\}_{k\in \mathcal{K}}$ and $\{\hat{P}_k^{sh}\}_{k\in \mathcal{K}}$. Then, 
each owner of a \ac{sh} will pay
\begin{equation}
    \hat{B}_{\mathcal{K}}^i = \sum_{k\in \mathcal{K}} \gamma^i_k \Delta t_s \hat{P}_k^i,
\end{equation}
while the \ac{rec} will receive
\begin{equation}
    \hat{D}_{\mathcal{K}} = \sum_{k\in \mathcal{K}}  \gamma_{sh} \Delta t_s \hat{P}_k^{sh} + \gamma^e_k \Delta t_s P_k^{pv}.
\end{equation}
This amount must be divided among the \ac{rec} members according to some \textit{fair} criterion. 

Actually, the discounted bill paid by the \ac{rec} members will be
\begin{equation}
    \hat{B}_{\mathcal{K}}^{D,i} = \sum_{k\in \mathcal{K}} \gamma^i_k \Delta t_s \hat{P}_k^i - \hat{D}^i_{\mathcal{K}},
\end{equation}
where $\hat{D}^i_{\mathcal{K}}$ is the discount allocated to the $i$-th member, which must be such that $\sum_{i=1}^N \hat{D}^i_{\mathcal{K}} = \hat{D}_{\mathcal{K}}$. A simple allocation policy could be 
\begin{equation}
\hat{D}^i_{\mathcal{K}}=\hat{D}_{\mathcal{K}}/N. 
\end{equation}
However, such a policy is unfair within the proposed scenario since \ac{rec} members are granting the control of their smart appliances to the \ac{cm}. Differently, each member could independently control its own devices without regard to the community discount and minimize 
\begin{equation}\label{eq:memb-cost-funct}
   J^i =  \sum_{k\in \mathcal{K}} \gamma^i_k\cdot \Delta t_s P_k^i. 
\end{equation}
By indicating with $\{\tilde{P}^i_k\}_{k\in\mathcal{K}}$ the power demand profiles obtained after the minimization of \eqref{eq:memb-cost-funct} and the \ac{rec} operation, each owner of a \ac{sh} will pay
\begin{equation}\label{eq:ubills}
    \tilde{B}_{\mathcal{K}}^i = \sum_{k\in \mathcal{K}} \gamma^i_k\cdot \Delta t_s \tilde{P}_k^i. 
\end{equation}
It is straightforward to prove that $\tilde{B}_{\mathcal{K}} \leq \hat{B}_{\mathcal{K}}$, \textit{i.e.} the bill minimized by the single members without taking into account the community discount is lower or equal to the one centrally minimized by the \ac{cm} considering the community discount. Consequently, without granting control of the its smart appliances to the \ac{cm} the $i$-th member experiences the economical loss
\begin{equation}\label{eq:loss}
    L_{\mathcal{K}}^i = \hat{B}_{\mathcal{K}}^i - \tilde{B}_{\mathcal{K}}^i.
\end{equation}
In fairness, this loss must be compensated using the community discount. Therefore, we propose the following discount allocation
\begin{equation}\label{eq:discount-allocation}
    \hat{D}^i_{\mathcal{K}} = L_{\mathcal{K}}^i + \frac{\hat{D}_{\mathcal{K}}-\sum_{i=1}^N L_{\mathcal{K}}^i}{N},
\end{equation}
where the total community discount is firstly used to offset the loss of members' bills and the remaining amount is then distributed equally.

After having introduced the \acp{sh}' models in Section~\ref{sec:models}, in Section~\ref{sec:algorithm} we propose a \ac{cma} composed by: 
\begin{itemize}
    \item[i)] the \ac{coa} that minimizes cost function \eqref{eq:cer-cost-function};
    \item[ii)] the \ac{moa} that minimizes the undiscounted members' bills \eqref{eq:memb-cost-funct}, in order to quantify losses \eqref{eq:loss} and implement the discount allocation policy \eqref{eq:discount-allocation}.
\end{itemize}    
%%%%%%%%%%%%%%%%%%%%%%%%%%%%%%%%%%%%%%%%%%%%%%%%%%%%%%%%%%%%%%%%
\section{Smart Homes Models}\label{sec:models}
%%%%%%%%%%%%%%%%%%%%%%%%%%%%%%%%%%%%%%%%%%%%%%%%%%%%%%%%%%%%%%%%
Each \ac{sh} is supposed to be equipped with the following of appliances
\begin{itemize}
    \item one \acp{acs} by which the user regulates the house internal temperature;
    \item two \acp{pa}, typically made up by a dishwasher and a washing machine;
    \item one \ac{pevrs};
    \item various \acp{ul}, such as refrigerators, TVs, computers, oven, air-dryer etc..
\end{itemize}

In this paper, \ac{acs}, \acp{pa} and \ac{pevrs} are smart appliances, that can be controlled remotely by the \ac{cm}. For these devices, users indicate their scheduling requirements in terms of start and ultimate end time. This information is represented by the binary parameters $U^{h,i}_k$, $U^{c,i}_k$, $U^{a,i}_k$, $U^{p,i}_k$; in particular, for the \ac{acs} are defined heating ($h$) and cooling ($c$) mode, for each of the two \acp{pa} ($a$) and for the \ac{pevrs} ($p$), respectively. Such binary quantities are set to 1 in the time interval indicated by the user, and to 0 when the devices are forced to be switched off. The idea in this paper is that users are not asked to declare in advance their scheduling requirements (\textit{e.g.}, at the beginning of the day for the next 24 hours), but, at any time step $k=k_1$, they can switch on any device and declare the ultimate end time $k_2>k_1$ in order to apply a easier approach to home automation.

In the next, we provide the models adopted for each of the above listed devices.

\subsection{Air-Conditioning System}
Let $\theta^i$ [°C] be the internal temperature of the $i$-th \ac{sh}. Its time dynamics is represented by a simplified one room equivalent model:
\begin{equation}\label{eq:temperature}
        \theta^i_{k+1} = \alpha^i \theta^i_k + \beta^i R^i \left( \eta_h^i P^{h,i}_k - \eta_c^i P_k^{c,i} \right) + \beta^i \theta^{ex}_k,
\end{equation}
where: $R^i$ [°C/kW] is the thermal resistance of the house walls; $\alpha^i=\exp(-\Delta t_s/R^iC^i)$, with $C^i$ [kWh/°C] being the thermal capacitance of the masses within the house; $\beta^i=1-\alpha^i$; $P^{h,i}$ and $P^{c,i}$ are the electrical powers consumed by the \ac{acs} to heat and cool, respectively; $\eta_h^i$ and $\eta^i_c$ are the heating and cooling \acp{cop}, respectively; $\theta^{ex}$ [°C] is the external ambient temperature (assumed to be the same for all the \acp{sh}). 

Obviously, the \ac{acs} cannot simultaneously work on heating and cooling mode. In this work we suppose that the user decides whether the \ac{acs} is on and in which mode. As mentioned before, to model the users requirements, binary parameters $U_k^{h,i}$ and $U_k^{c,i}$ are defined. The \ac{acs} is supposed to be inverter-driven, meaning that heating or cooling powers can be modulated with continuity from zero to the nominal powers $P^{h,i}_{nom}$ and $P^{c,i}_{nom}$ (usually different each others). Thus, the following constraints must be satisfied
\begin{align}
    0&\leq P_k^{h,i}\leq U_k^{h,i} P_{nom}^{h,i}, \\
    0&\leq P_k^{c,i}\leq U_k^{c,i} P_{nom}^{c,i}.
\end{align}

When working in heating (cooling) mode, the \ac{acs} objective is to keep the internal temperature higher (lower) than the set-point $\theta^i_{sp}$. Therefore, the following constraint must be satisfied
\begin{equation}\label{eq:temperature_objective}
U_k^{h,i}\theta^i_{sp}\leq (U_k^{h,i}+U_k^{c,i}) \theta^i_k \leq U_c^{h,i}\theta^i_{sp}
\end{equation}
Note that when the \ac{acs} is off $U_k^{h,i}+U_k^{c,i}=0$, and constraint \eqref{eq:temperature_objective} is trivially satisfied.

%------------------------------------------
\subsection{Phases-based Appliances}
Home appliances, such as dishwashers and washing machines, usually execute a program selected by the user, consisting of a set of consecutive phases. Each phase has a definite power level. Therefore, the selected program is associated to a specific power profile. 

In this paper, we assume that such a profile is known, at least since the program is selected. Moreover, as mentioned before, we suppose that the user switches on the \ac{pa} at time $k=k_1$ and declares that the entire program must end up to time $k_2\geq M^i-k_1$, where $M^i$ is the duration (\# of time steps) of the selected program. Thus, $U^{a,i}_k=1$ for all $k\in[k_1,k_2]$ and $U^{a,i}_k=0$ for all $k\notin[k_1,k_2]$.

To model the operation of the \ac{pa}, we define the following binary variables
\begin{itemize}
\item $\delta_k^{j,i}$, $j=1,2,\ldots,M^i$, equal to 1 if the $j$-th phase is taking place at time $k$;
\item $s_k^{i}$, equal to 1 if the entire program has been completed at time $k-1$;
\end{itemize}   
and the power profile $\{\hat{P}_j^i\}_{j=1}^{M^i}$ associated to the selected program.

At the starting time $k_1$, the binary variables are initialized as follows
\begin{align}
    &s_{k_1}^i = 0, \label{eq:aop6}\\
    &\delta_{k_1}^{j,i} = 0 \quad \forall j\in[2,M^i].   \label{eq:aop7}  
\end{align}
Then, power $P^{a,i}_k$ consumed by the \ac{pa} at time $k\geq k_1$ is given by
\begin{equation}\label{eq:Pptot}
    P^{a,i}_k = \sum_{j=1}^{M^i} \delta_k^{j,i} \hat{P}_j^i
\end{equation}
and the following constraints must be satisfied
\begin{align}
    &\sum_{j=1}^{M^i} \delta_k^{j,i} \leq U^{a,i}_k & &\forall k\geq k_1,\label{eq:aop1} \\
    &\sum_{k \geq k_1}\sum_{j=1}^{M^i} \delta_k^{j,i} = M^is_k^{i}, \label{eq:aop1bis} \\
    &\delta_{k+1}^{j+1,i} = \delta_k^{j,i},  & &\forall \ j\in[1,M^i-1], k\geq k_1,\label{eq:aop2}\\
    &\delta_k^{M^i,i}=s_{k+1}^i, & &\forall  k\geq k_1,\label{eq:aop3}\\
    &s_k^i-s^i_{k+1}\leq 0, & &\forall k\geq k_1,\label{eq:aop4}\\
    &\delta_k^{j,i} \leq 1-s_k^i, & &\forall \ j\in[1,M^i],k\geq k_1. \label{eq:aop5}  
\end{align}
Constraint \eqref{eq:aop1} imposes that the \ac{pa} can work only if desired by the user and that, at any time step $k$, just one phase can take place; constraint \eqref{eq:aop1bis} imposes that the program must be completed at the end of the indicated time interval; constraint \eqref{eq:aop2} imposes the order of the phases; constraints \eqref{eq:aop3}-\eqref{eq:aop4} allow $s_k^{i}$ to become equal to 1 when the entire program is ended, according to its definition; constraint \eqref{eq:aop5} avoids the program re-start after its termination.  

Obviously, if the $i$-th \ac{sh} is equipped with two (or more) \acp{pa}, all the above defined variables and parameters as well as model \eqref{eq:aop6}--\eqref{eq:aop5} should be defined for each \ac{pa}.

%----------------------------------
\subsection{PEV Recharging Station}
For \ac{pevrs}, we suppose that at time step $k=k_1$ the $k$-th user connects its \ac{pev} and declares that the recharge must finish up to time $k_2$. Thus, $U^{p,i}_k=1$ for all $k\in[k_1,k_2]$ and $U^{a,i}_k=0$ for all $k\notin[k_1,k_2]$. Moreover, at time $k_1$, following data are available 
\begin{itemize}
    \item \ac{pev} battery capacity $E_b^i$ [kWh];
    \item \ac{pev} battery recharging efficiency $\eta_b^i$;
    \item \ac{pev} battery \ac{soc} $SoC_{k_1}$ [p.u.].
\end{itemize}
 
 We suppose that the \ac{pevrs} is inverter-driven so that the recharging power $P^{p,i}_k$ [kW] can be modified with continuity from zero to to the nominal value of the charging facility $P^{p,i}_{nom}$. 

Given these definitions, the \ac{pevrs} operation can be modelled by the following equations
\begin{equation}
    \Delta t_s \sum_{k\geq k_1} U^{p,i}_k \eta_b^i P^{p,i}_k = E_b^i(1-SoC_{k_1}), \label{eq:pev_con1}
\end{equation}
\begin{equation}
    0\leq P^{p,i}_k \leq P^{p,i}_{nom}U_k^{p,i}. \label{eq:pev_con2}
\end{equation}

Note that constraint \eqref{eq:pev_con1} imposes that the \ac{pev} battery must be fully recharged up to time $k_2$. This means that it can be satisfied only if 
\begin{equation}
    \Delta t_s \sum_{k\geq k_1} U^{p,i}_k \eta_b^i P^{p,i}_{nom} \geq E_b^i(1-SoC_{k_1}), \label{eq:pev_limit}
\end{equation}
\textit{i.e.} the declared recharging time interval is enough to fully recharge the \ac{pev} battery, assuming to recharge at the rated power $P^{p,i}_{nom}$. Without loss of generality, in this paper we assume that \eqref{eq:pev_limit} is always satisfied. Differently, there are two possibilities: the first one is to relax constraint \eqref{eq:pev_con1} and recharge as much as possible; the second one is to compute the maximal \ac{soc} reachable within the declared time interval and use it to substitute the unitary quantity in the right member of \eqref{eq:pev_con1}.

%-------------------------------
\subsection{Smart Home System}
The total power consumed by the $i$-th \ac{sh} at time $k$ is finally given by
\begin{equation} \label{eq:balance}
    P^i_k = P^{h,i}_k+P^{c,i}_k+P^{a,i}_k+P^{p,i}_k+P^{u,i}_k,
\end{equation}
where $P^{u,i}_k$ is the power consumed by the \acp{ul}. Total power  $P^i_k$ cannot overtake the rated power $P^i_{max}$ established by the contract with the energy retailer, that is
\begin{equation}\label{eq:Pimax}
    P^i_k \leq P^i_{max}, \quad \forall k. 
\end{equation}
%%%%%%%%%%%%%%%%%%%%%%%%%%%%%%%%%%%%%%%%%%%%%%%%%%%%%%%%%%%%%%%%%
\section{The Community Management Algorithm}\label{sec:algorithm}
%%%%%%%%%%%%%%%%%%%%%%%%%%%%%%%%%%%%%%%%%%%%%%%%%%%%%%%%%%%%%%%%%
The \ac{cm} receives at any time step $k$ 
\begin{itemize}
    \item the updated measurements of the \ac{pv} power generation $P^{pv}_k$;
    \item the power consumption from each \ac{sh} $P^i_k$,  
    the \acp{sh} state variables (\textit{i.e} $\theta^i_k$);
    \item the users' time scheduling for the controlled devices (\textit{i.e.}, $U^{h,i}_k$, $U^{c,i}_k$, $U^{a,i}_k$, $U^{p,i}_k$);
    \item the control specifications (\textit{i.e.}, $\theta_{sp}^i$, $SoC_{k_1}$, selected program of the \acp{pa}). 
\end{itemize}

Moreover, the \ac{cm} is supposed to know all modeling parameters and to receive, at any time step $k$, the forecasted profiles of the \ac{pv} power generation $\{\hat{P}_{\tau}^{pv}\}_{\tau=k}^{T-1}$, of the \acp{ul} $\{\hat{P}_{\tau}^{u,i}\}_{\tau=k}^{T-1}$, for all $i=1,2,\ldots,N$, and of the external ambient temperature $\{\hat{\theta}^{ex}_\tau\}_{\tau=k}^{T-1}$. Here, $T$ is the control time horizon, expressed in terms of number of time steps.

The control variables decided by the \ac{cm} are the powers absorbed by the smart appliances $P^{h,i}_k$, $P^{c,i}_k$, $P_k^{a,i}$, $P_k^{p,i}$ for all $i=1,2,\ldots,N$. For simplicity, in the following we will assume that all variables are stacked in the control vectors $u_k^i$, in turn stacked in the overall control vector $u_k$.

As introduced in Section~\ref{sec:scenario}, our \ac{cma} is composed by two optimization algorithms, the \ac{coa} and the \ac{moa}, which have the objective of minimizing cost function \eqref{eq:cer-cost-function} and \eqref{eq:memb-cost-funct} for all $i=1,2,\ldots,N$, respectively. 

To perform these optimizations, \ac{mpc} is adopted \cite{Morari:1988}. In brief, \ac{mpc} consists in repeating the following procedure each time step $k$:
1) to define a control optimization problem over the time interval $[k,k+T-1]$, based on updated measurements and forecasts; 2) to solve this optimization problem obtaining the optimal control vector trajectory $\{u^*_\tau\}_{\tau=k}^{k+T-1}$; 3) to apply the first element of this trajectory $u^*_k$. 

This procedure, based on the so called \textit{receding horizon principle}, is
well known to be robust with respect to modeling and measurements errors since decisions are dynamically adapted to updated data. Moreover, it is particularly suited for our control scenario, where users are free to dynamically change their requirements.

%---------------------------------------------
\subsection{Community Optimization Algorithm}\label{ssec:coa}
The \ac{coa} applies \ac{mpc} to the control optimization problem defined as follows:
\begin{equation}\label{eq:coa-cost-function}
\{u^*_\tau\}_{\tau=k}^{k+T-1} = \arg\min \sum_{\tau=k}^{k+T-1} \left[\sum_{i=1}^N (\gamma^i_\tau P_\tau^i) -  \gamma_{sh} P_\tau^{sh}\right]
\end{equation}
such that
\begin{align}
&P_\tau^{sh} \leq \hat{P}^{pv}_\tau, & &\forall \tau\in[k,k+T-1],\label{eq:shared-con1} \\
&P_\tau^{sh} \leq  \sum_{i=1}^N P_\tau^i, & &\forall \tau\in[k,k+T-1], \label{eq:shared-con2}
\end{align}
and constraints \eqref{eq:temperature}--\eqref{eq:pev_con2}, \eqref{eq:balance}--\eqref{eq:Pimax} 
are satisfied over the time horizon $[k,k+T-1]$ for all $i=1,2,\ldots,N$, and given the updated measurements and forecast trajectories.

Note that the definition \eqref{eq:shared} of $P^{sh}_k$  is included in the optimization problem through the two linear inequalities \eqref{eq:shared-con1} and \eqref{eq:shared-con2}. Indeed, in virtue of cost function \eqref{eq:coa-cost-function}, $P^{sh}_k$ will be maximized and thus leaded to be equal to the minimum between $\hat{P}^{pv}_k$ and $\sum_{i=1}^N P_k^i$.

%---------------------------------------------
\subsection{Member Optimization Algorithm} \label{ssec:moa}
The \ac{moa} applies \ac{mpc} to the control optimization problems defined as follows, for all $i=1,2,\ldots,N$:
\begin{equation}\label{eq:moa-cost-function}
\{{u^{i}_\tau}^*\}_{\tau=k}^{k+T-1} = \arg\min \sum_{\tau=k}^{k+T-1} \sum_{i=1}^N \gamma^i_\tau P_\tau^i
\end{equation}
such that constraints \eqref{eq:temperature}--\eqref{eq:pev_con2}, \eqref{eq:balance}--\eqref{eq:Pimax} are satisfied over the time horizon $[k,k+T-1]$ and given the updated measurements and forecast trajectories.

%\vspace{15pt}
Both the \ac{coa} and \ac{moa} optimization problems are linear mixed-integer and can be solved by any suitable solver. It is worth noting that while the \ac{coa} is centralized, the \ac{moa} can be run in parallel for each member, so possibly even within a distributed computational framework. 

%%%%%%%%%%%%%%%%%%%%%%%%%%%%%%%%%%%%%%%%%%%%%%%%%%%%%
\section{Simulations and Results}\label{sec:results}
\begin{figure}[t]
	\centering
    \includegraphics[width=1\columnwidth]{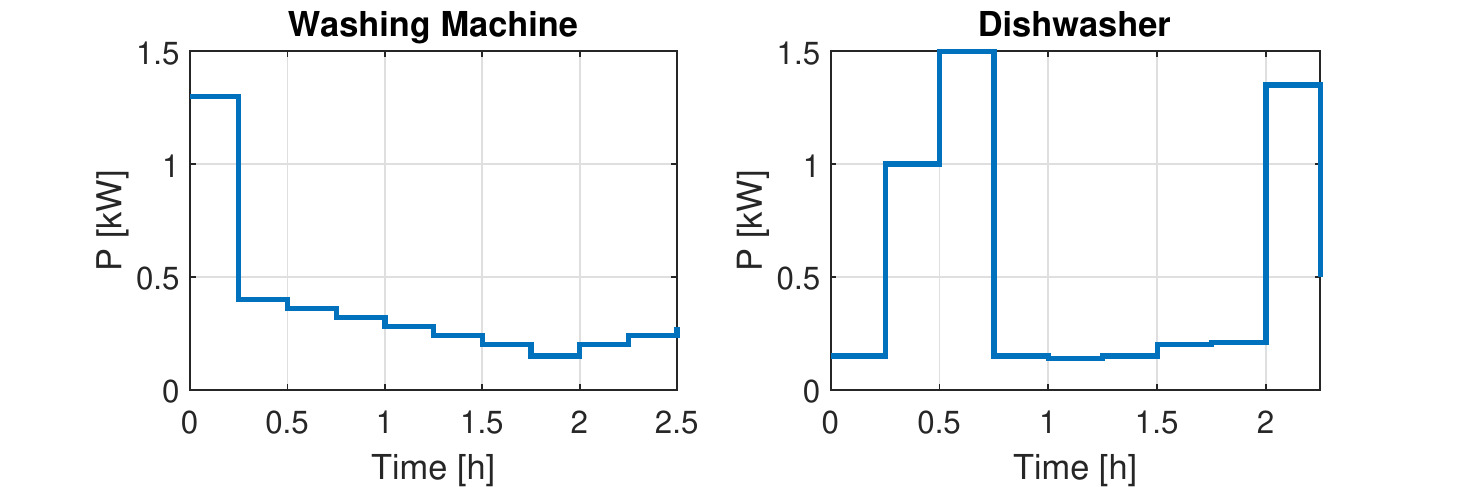}
\caption{\acp{pa}' average program power profiles ($\{\hat{P}_j^i\}_{j=1}^{M^i}$, $M^i=11$ for washing machines and $M^i=10$ for dishwashers).}
\label{fig:abps}
\end{figure}

%%%%%%%%%%%%%%%%%%%%%%%%%%%%%%%%%%%%%%%%%%%%%%%%%%%%%
To test the effectiveness of the proposed method, a \ac{rec} with $N=10$ \acp{sh} is simulated. They share a 15~kWp \ac{pv} plant and their rated import power is $P_{nom}^i=6$~kW. Monitoring and control sampling time is $\Delta t_s = 0.25$~h (one quarter of hour). \acp{acs} and \acp{pevrs} parameters are randomly generated, varying from the average values reported in Table~\ref{tab:paramters}, up to the 20\%. Each \ac{sh} is equipped with one smart washing machine and one smart dishwasher. Their program profiles are randomly generated  varying from the average profiles in Fig.~\ref{fig:abps}, up to the 20\%. These profiles are extracted from a database of the University of Genova (Italy), collecting real measurements from four test flats. The same database is used to generate \acp{ul} power profiles.
\ac{pv} power generation and external ambient temperature are collected from the NASA POWER dataset\footnote{These data were obtained from the NASA Langley Research Center (LaRC) POWER Project funded through the NASA Earth Science/Applied Science Program.} referring to the area of Rome (Italy). 

\begin{table}[t]
\caption{ACSs and PEV-RSs average parameters}
\begin{center}
\begin{tabular}{lccc}
\hline
Parameter & Symbol & Value & Unit \\
\hline
\vspace{-3pt}\\
\multicolumn{4}{l}{\textbf{ACSs}} \\
 Thermal resistance & $R$ & 12.5 & °C/kW \\
 Thermal capacitance & $C$ & 0.1 &  kWh/°C \\
 Heating \ac{cop} & $\eta_h$ & 3.5 & - \\
 Cooling \ac{cop} & $\eta_c$ & 3 & - \\
 Heating rated power & $P_{nom}^h$& 1 & kW \\
 Cooling rated power & $P_{nom}^c$& 0.7 & kW \\
 Temperature set-point & $\theta_{sp}$ & 23 & °C \\
\vspace{-3pt}\\
\multicolumn{4}{l}{\textbf{PEV-RSs}} \\
Rated power & $P^p_{nom}$ & 3.6 & kW \\
Battery capacity & $E_b$ & 15 & kWh \\
Recharging efficiency & $\eta_b$ & 0.95 & - \\
\hline
\end{tabular}
\label{tab:paramters}
\end{center}
\end{table}

\begin{figure}[t]
	\centering
    \includegraphics[width=1\columnwidth]{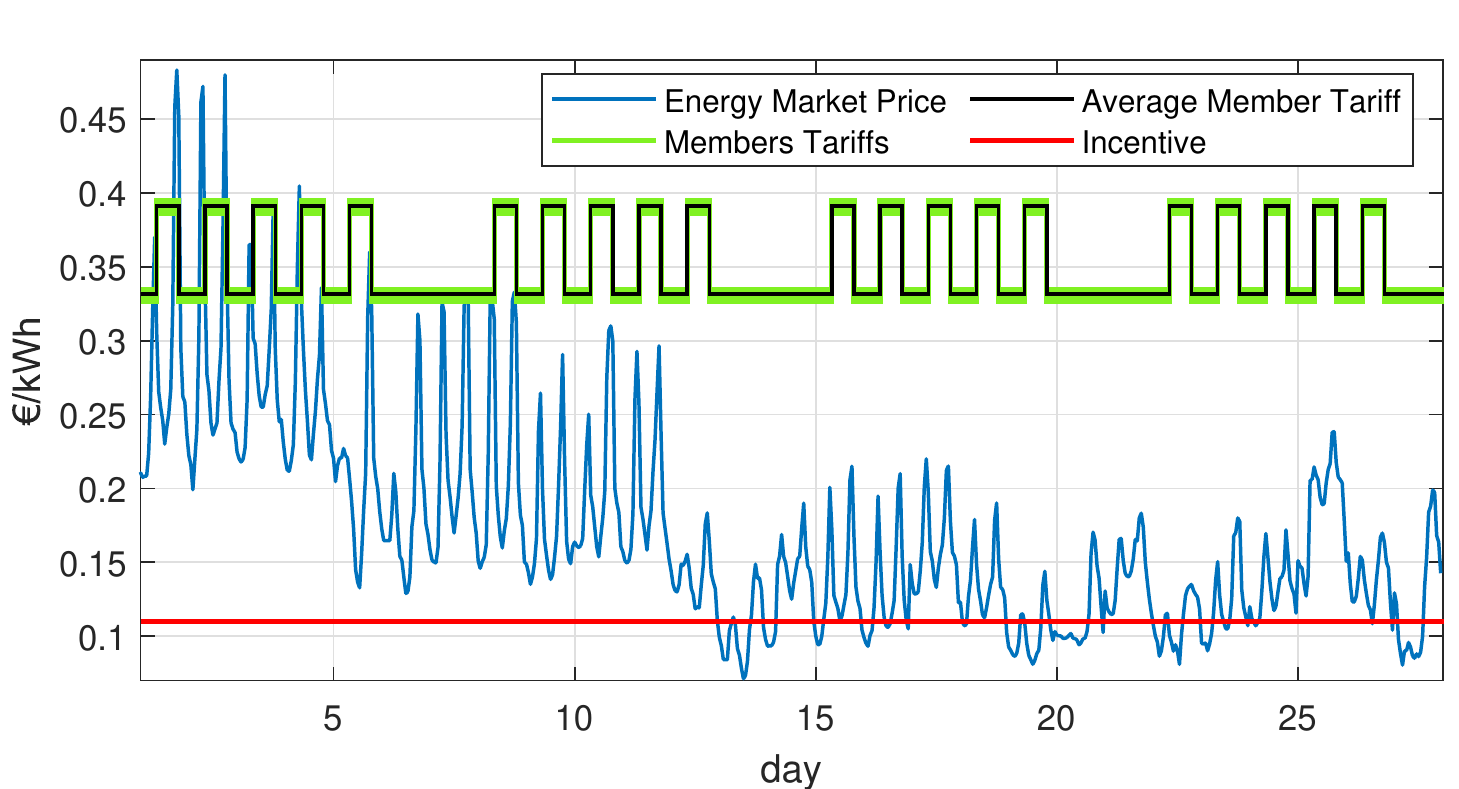}
\caption{Energy market price, members' tariffs and incentive.}
\label{fig:prices}
\end{figure}

The presented simulation results cover 28 days from June 13 to July 10, 2022. Figure~\ref{fig:prices} shows: the members' energy use tariffs ($\gamma^i_k$), the energy market price ($\gamma^e_k$), payed to the \ac{rec} for the \ac{pv} generation, and the incentive paid to the \ac{rec} proportionally to the \ac{se} ($\gamma_{sh}$), fixed to 0.11~\euro/kWh. Members' tariffs are based on an average calculated over a set of commercial proposals from Italian retailers. As shown in  Figure~\ref{fig:prices}, there are two price levels during weekdays, the highest from 8 a.m. to 7 p.m., and the lowest during the remaining hours of the day and on weekends.


\begin{table}[t]
\caption{Smart appliances on/off time intervals.}
\begin{center}
\begin{tabular}{lll}
\hline
\vspace{-3pt}\\
 \ac{acs} & ON & 6 a.m. - 9 a.m.\\
 & OFF & 5 p.m. - 8 p.m. \\
 \vspace{-3pt}\\
 \ac{pevrs} & ON & 4 p.m. - 10 p.m.\\
 & OFF & 6 p.m. - 12 p.m. (day after) \\
  \vspace{-3pt}\\
 {Washing Machine} & ON & 6 p.m. - 11 p.m.\\
 & OFF & 4 p.m. - 6 p.m. (day after) \\
  \vspace{-3pt}\\
  {Dishwasher} & ON & 6 p.m. - 11 p.m.\\
 & OFF & 4 p.m. - 6 p.m. (day after) \\
 \hline
\end{tabular}
\label{tab:user-requirements}
\end{center}
\end{table}

Smart appliances on/off timing are generated randomly within time intervals reported in Table~\ref{tab:user-requirements}. \acp{acs} are in cooling mode. When a \ac{pevrs} is switched on, the battery initial \ac{soc} ($SoC_{k_1}^i$) is randomly set. 

\ac{pv} generation and external ambient temperature forecasts are generated by randomly corrupting real data, with a maximum error of $10\%$. Whereas, the \ac{ul} power profiles are predicted using an average over the preceding seven days. It is worth noting that the method used to perform forecasts is beyond the scope of this paper. 

The \ac{cma} described in Section~\ref{sec:algorithm} has been implemented using the AMPL programming language \cite{AMPL} integrated into the MATLAB simulation platform. Optimizations are solved by the CPLEX solver. Simulations are performed using a Intel(R) Core(TM) i7 CPU@1.30 GHz. The average computational time for one control step (to be executed at each quarter of hour in a real application) results to be about 2~seconds.

\begin{figure}[t]
	\centering
    \includegraphics[width=1\columnwidth]{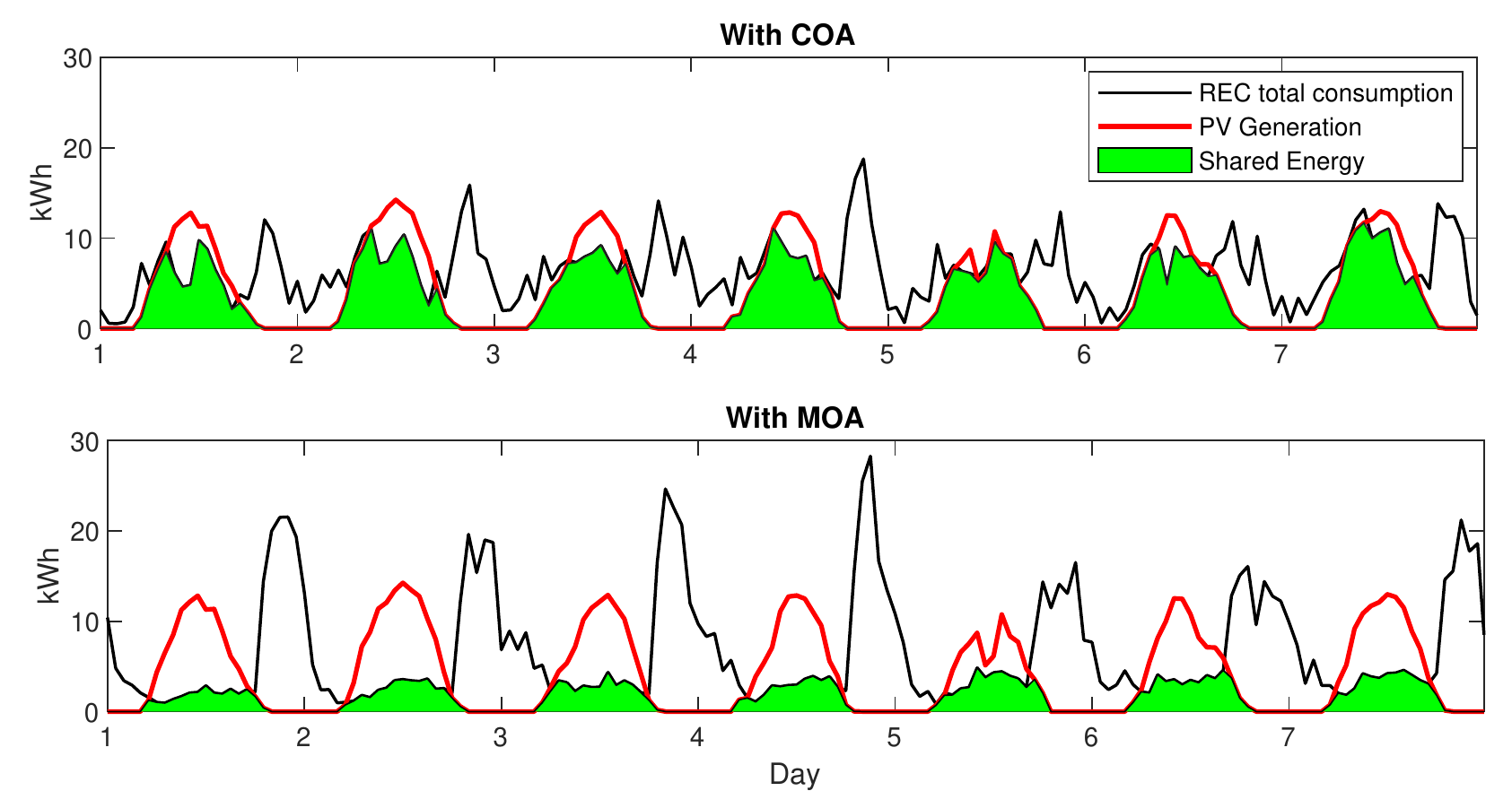}
\caption{\ac{pv} generation, total \ac{rec} consumption and \ac{se} during the first week of simulation.}
\label{fig:pvs}
\end{figure}
\begin{figure}[h]
	\centering
    \includegraphics[width=1\columnwidth]{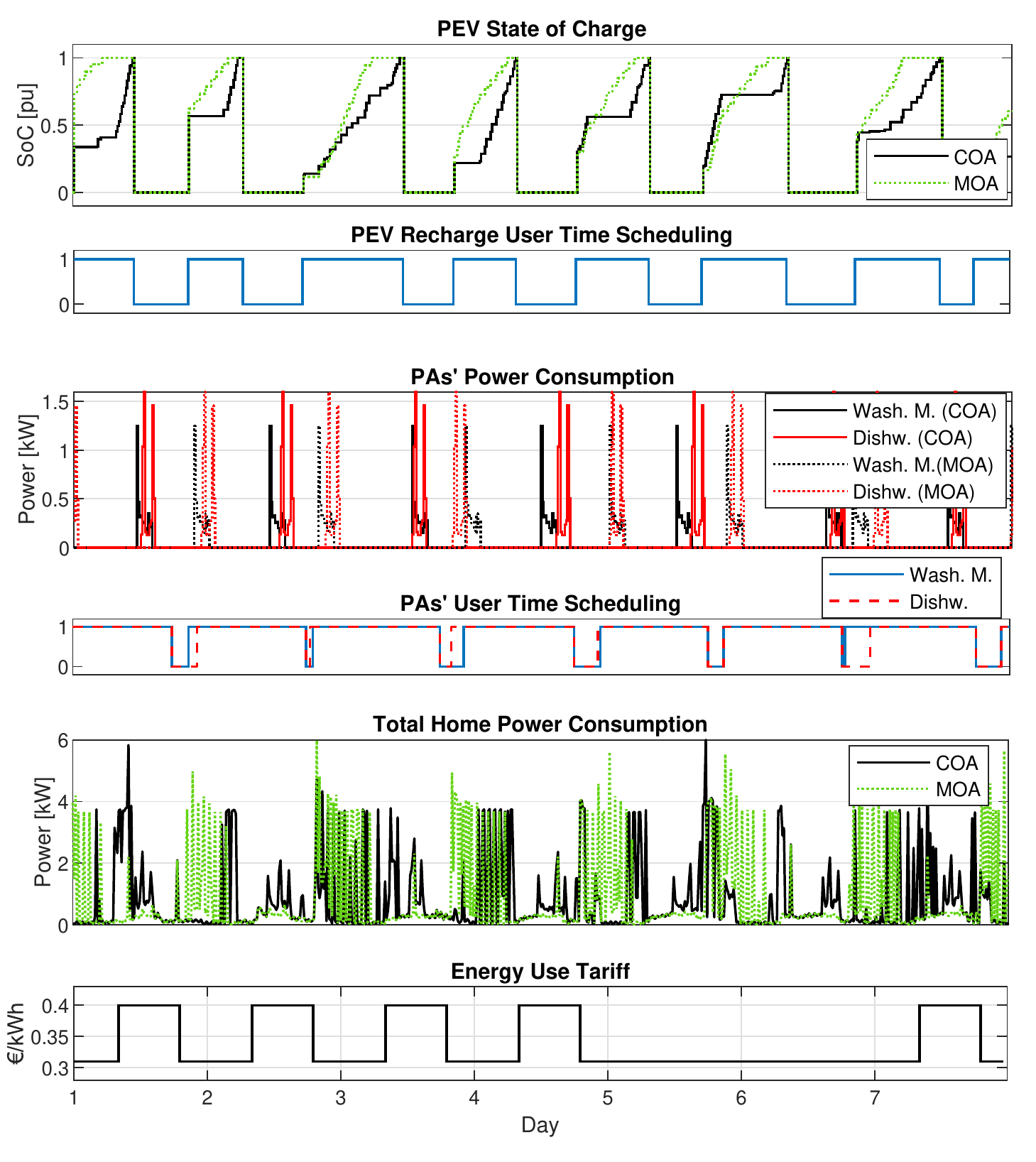}
\caption{Results for SH 1 during the first week of simulation.}
\label{fig:powers}
\end{figure}

Figures~\ref{fig:pvs} and \ref{fig:powers} show a sketch of the results obtained within the first week of simulation. In Fig.~\ref{fig:pvs}, we can observe the aggregated \ac{rec} energy demand, the \ac{pv} generation profile and the consequent \ac{se}: in the top figure, we have the profiles obtained with the \ac{coa} (Section~\ref{ssec:coa}), which coordinates the members' consumption to minimize their bills, taking into account the discount coming from the \ac{se}; in the bottom figure, we have the profiles realized with the \ac{moa} (Section~\ref{ssec:moa}), which minimizes the bills of single members, without considering the \ac{se} discount. By comparing the two figures, it clearly appears that \ac{se} is significantly increased by the \ac{coa}, leading \ac{rec} members to reduce their consumption at night and increase it during daylight hours. 

In Fig.~\ref{fig:powers}, we can observe how the \ac{coa} operates on a \ac{sh}. Given the time flexibility granted by the user for \ac{pevrs} and \acp{pa}, and the time-varying energy use tariff (bottom figure), the \ac{moa} obviously decides to recharge the \ac{pev} and perform \acp{pa} programs when tariff is lower. Differently, when possible, the \ac{coa} moves \ac{pev} recharge and \acp{pa}' program executions during daylights, when \ac{pv} generation is available.

At the end of the four weeks of simulation, the total \ac{se} obtained with the \ac{coa} is about 2325~kWh, versus 1105~kWh realized with the \ac{moa}. Thus, thanks to the proposed \ac{coa}, the \ac{se} is more than doubled, as is the resulting discount, which increases from 122~\euro \ to 255~\euro.

\begin{figure}[t]
	\centering
    \includegraphics[width=1\columnwidth]{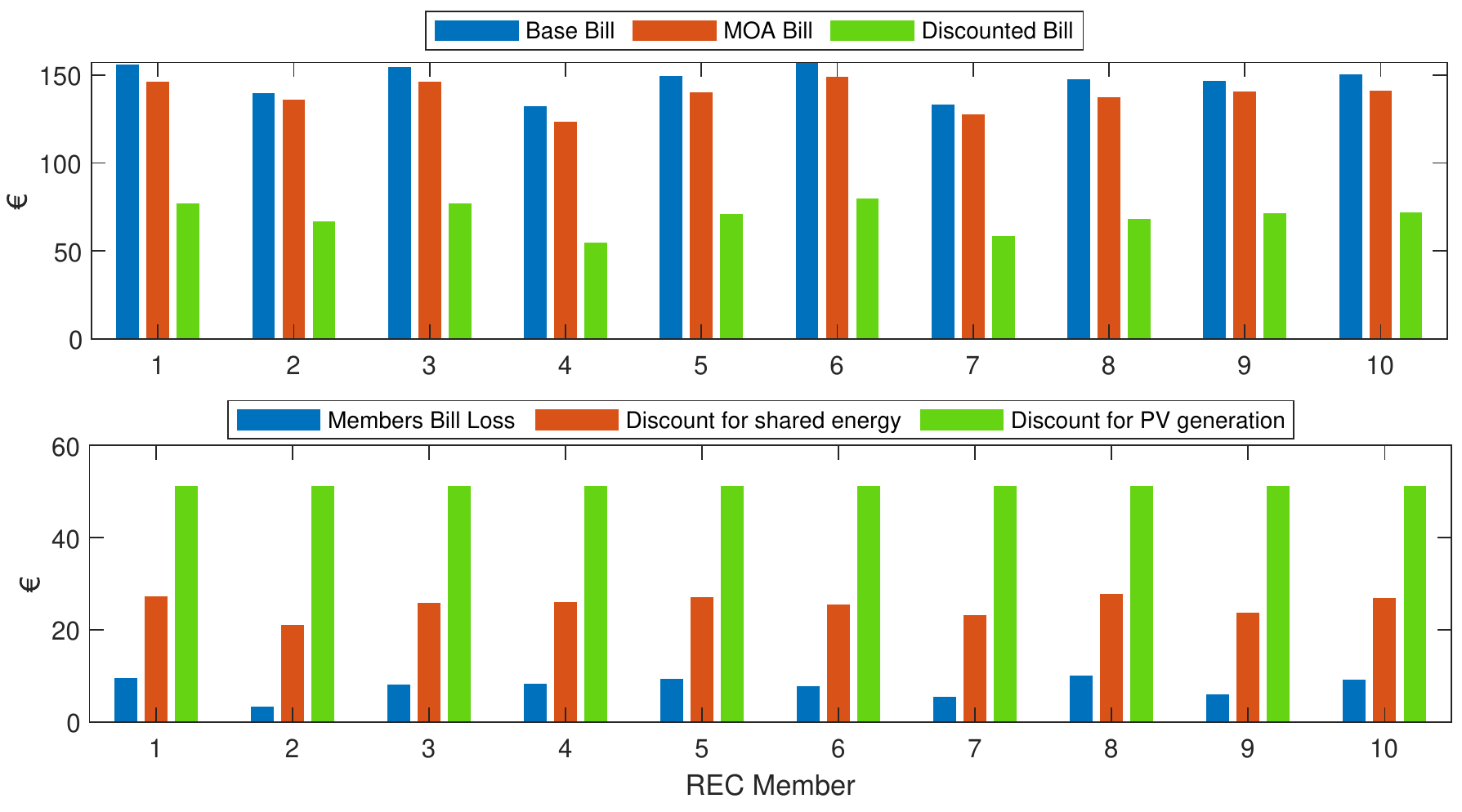}
\caption{Members bills.}
\label{fig:bills}
\end{figure}

Figure~\ref{fig:bills} shows the resulting bills of the \ac{rec} members. In the top figure, we have the undiscounted bills, compared with the ones obtained by applying the MOA with and without the final discount. As expected and discussed in Section~\ref{sec:scenario}, \ac{moa} bills are lower than the undiscounted ones. Losses amounts ($L^i_{\mathcal{K}}$) are reported in the bottom figure, where they are compared with the two portions of the discount, coming from the \ac{se} and from selling the \ac{pv} energy. Discount allocation policy \eqref{eq:discount-allocation} is applied to the total discounts, but in the figure it is done on the \ac{se} ones. We remark that, in any case, even just the \ac{se} discount, is enough to compensate members' bill losses. We also note that, finally, discounted bills result to be approximately reduced by half. 

Moreover, we cannot avoid observing that the discount from \ac{se} is significantly lower than the one due to energy selling. This can be explained by observing in Fig.~\ref{fig:prices} that the energy market price is on average much higher than the value of the \ac{se} incentive. It is worth remaking that the order of magnitude of the latter was designed and defined in Italy 
during 2020, when the average electrical energy market price was around 0.05~\euro/kWh.

%\begin{figure}[t]
%	\centering
%    \includegraphics[width=1\columnwidth]{comparison.eps}
%\caption{Impact of the \ac{coa} on \ac{se}.}
%\label{fig:comparison}
%\end{figure}

%%%%%%%%%%%%%%%%%%%%%%%%%%%%%%%%%%%%%%%%%%%%%%%%%%%
\section{Conclusions}\label{sec:conclusions}
%%%%%%%%%%%%%%%%%%%%%%%%%%%%%%%%%%%%%%%%%%%%%%%%%%%
In this paper, we have proposed a management algorithm for a \ac{rec} composed by a set of \acp{sh} sharing a \ac{pv} plant, complaint with the Italian regulatory framework. The control strategy allows \ac{rec} members' consumption profiles to be coordinately modified to maximize the discount achieved by the incentive provided to the community for sharing renewable energy. Results show that the proposed method is effective and that, within the proposed scheme, joining the \ac{rec} and granting control and time flexibility to a community management algorithm can allow users to achieve significant savings in their electricity bills.  

Future works will be dedicated to several different developments of the approach such as studying a distributed version of the community optimization algorithm or coupling it with a data-driven identifier for the houses thermal dynamics and for the \acp{pa} programs profiles.     
 
\bibliographystyle{IEEEtran}
\bibliography{biblio.bib}

\begin{thebibliography}{10}
\providecommand{\url}[1]{#1}
\csname url@samestyle\endcsname
\providecommand{\newblock}{\relax}
\providecommand{\bibinfo}[2]{#2}
\providecommand{\BIBentrySTDinterwordspacing}{\spaceskip=0pt\relax}
\providecommand{\BIBentryALTinterwordstretchfactor}{4}
\providecommand{\BIBentryALTinterwordspacing}{\spaceskip=\fontdimen2\font plus
\BIBentryALTinterwordstretchfactor\fontdimen3\font minus
  \fontdimen4\font\relax}
\providecommand{\BIBforeignlanguage}[2]{{%
\expandafter\ifx\csname l@#1\endcsname\relax
\typeout{** WARNING: IEEEtran.bst: No hyphenation pattern has been}%
\typeout{** loaded for the language `#1'. Using the pattern for}%
\typeout{** the default language instead.}%
\else
\language=\csname l@#1\endcsname
\fi
#2}}
\providecommand{\BIBdecl}{\relax}
\BIBdecl

\bibitem{REDII}
{European Union}, ``{Directive (EU) 2018/2001 of the European Parliament and of
  the Council of 11 December 2018 on the promotion of the use of energy from
  renewable sources.}'' \emph{Official Journal of the European Union}, vol.~61,
  2018.

\bibitem{Faia:2021}
R.~Faia, J.~Soares, T.~Pinto, F.~Lezama, Z.~Vale, and J.~M. Corchado, ``Optimal
  model for local energy community scheduling considering peer to peer
  electricity transactions,'' \emph{IEEE Access}, vol.~9, pp. 12\,420--12\,430,
  2021.

\bibitem{Pereira:2022}
H.~Pereira, L.~Gomes, and Z.~Vale, ``Peer-to-peer energy trading optimization
  in energy communities using multi-agent deep reinforcement learning,'' in
  \emph{Proceedings of the Energy Informatics Academy Conference 2022}, vol.~5,
  no.~44, 2022.

\bibitem{Feng:2020}
C.~Feng, F.~Wen, S.~You, Z.~Li, F.~Shahnia, and M.~Shahidehpour, ``Coalitional
  game-based transactive energy management in local energy communities,''
  \emph{IEEE Transactions on Power Systems}, vol.~35, no.~3, pp. 1729--1740,
  2020.

\bibitem{Safdarian:2021}
A.~Safdarian, P.~H. Divshali, M.~Baranauskas, A.~Keski-Koukkari, and
  A.~Kulmala, ``Coalitional game theory based value sharing in energy
  communities,'' \emph{IEEE Access}, vol.~9, pp. 78\,266--78\,275, 2021.

\bibitem{VanCutsem:2020}
O.~{Van Cutsem}, D.~{Ho Dac}, P.~Boudou, and M.~Kayal, ``Cooperative energy
  management of a community of smart-buildings: A blockchain approach,''
  \emph{International Journal of Electrical Power and Energy Systems}, vol.
  117, p. 105643, 2020.

\bibitem{Hafiz:2019}
F.~Hafiz, A.~{Rodrigo de Queiroz}, P.~Fajri, and I.~Husain, ``Energy management
  and optimal storage sizing for a shared community: A multi-stage stochastic
  programming approach,'' \emph{Applied Energy}, vol. 236, pp. 42--54, 2019.

\bibitem{DiLorenzo:2021}
G.~{Di Lorenzo}, S.~Rotondo, R.~Araneo, G.~Petrone, and L.~Martirano,
  ``Innovative power-sharing model for buildings and energy communities,''
  \emph{Renewable Energy}, vol. 172, pp. 1087--1102, 2021.

\bibitem{Zatti:2021}
M.~Zatti, M.~Moncecchi, M.~Gabba, A.~Chiesa, F.~Bovera, and M.~Merlo, ``Energy
  communities design optimization in the italian framework,'' \emph{Applied
  Sciences}, vol.~11, no.~11, 2021.

\bibitem{Moncecchi:2020}
M.~Moncecchi, S.~Meneghello, and M.~Merlo, ``A game theoretic approach for
  energy sharing in the italian renewable energy communities,'' \emph{Applied
  Sciences}, vol.~10, no.~22, 2020.

\bibitem{Ghiani:2019}
E.~Ghiani, A.~Giordano, A.~Nieddu, L.~Rosetti, and F.~Pilo, ``Planning of a
  smart local energy community: The case of berchidda municipality ({Italy}),''
  \emph{Energies}, vol.~12, no.~24, 2019.

\bibitem{Galici:2021}
M.~Galici, M.~Mureddu, E.~Ghiani, G.~Celli, F.~Pilo, P.~Porcu, and B.~Canetto,
  ``Energy blockchain for public energy communities,'' \emph{Applied Sciences},
  vol.~11, no.~8, 2021.

\bibitem{Conte:2022a}
F.~Conte, F.~D’Antoni, G.~Natrella, and M.~Merone, ``A new hybrid ai optimal
  management method for renewable energy communities,'' \emph{Energy and AI},
  vol.~10, p. 100197, 2022.

\bibitem{Conte:2022b}
F.~Conte, G.~Mosaico, G.~Natrella, M.~Saviozzi, and F.~R. Bianchi, ``Optimal
  management of renewable generation and uncertain demand with reverse fuel
  cells by stochastic model predictive control,'' in \emph{2022 17th
  International Conference on Probabilistic Methods Applied to Power Systems
  (PMAPS)}, 2022, pp. 1--6.

\bibitem{DiFazio:2022}
A.~R. Di~Fazio, A.~Losi, M.~Russo, F.~Cacace, F.~Conte, G.~Iannello,
  G.~Natrella, and M.~Saviozzi, ``Methods and tools for the management of
  renewable energy communities: the comer project,'' in \emph{2022 AEIT
  International Annual Conference (AEIT)}, 2022, pp. 1--6.

\bibitem{Morari:1988}
M.~Morari, C.~E. Garcia, and D.~M. Prett, ``Model predictive control: Theory
  and practice,'' \emph{IFAC Proceedings Volumes}, vol.~21, no.~4, pp. 1--12,
  1988, iFAC Workshop on Model Based Process Control, Atlanta, GA, USA, 13-14
  June.

\bibitem{AMPL}
{AMPL}, \url{https://ampl.com/}, accessed: 2023-01-12.

\end{thebibliography}
\end{document}